\documentclass[aps,prl,twocolumn,groupedaddress]{revtex4-1}

\usepackage{graphicx}
\usepackage{amssymb}
\usepackage{amsmath}

\begin{document}


\title{{\em Ab initio} many-body calculation of the $^7$Be($p$,$\gamma$)$^8$B radiative capture}


\author{Petr Navr{\'a}til$^{1,2}$, Robert Roth$^3$, and Sofia Quaglioni$^2$}
\affiliation{$^1$TRIUMF, 4004 Wesbrook Mall, Vancouver, BC V6T 2A3, Canada\\
$^2$Lawrence Livermore National Laboratory, P.O. Box 808, L-414, Livermore, CA 94551, USA\\
$^3$Institut f\"{u}r Kernphysik, Technische Universit\"{a}t Darmstadt, 64289 Darmstadt, Germany}

%
\date{\today}
\begin{abstract}
We apply the {\em ab initio} no-core shell model/resonating group method (NCSM/RGM) approach to calculate the cross section of the $^7$Be($p$,$\gamma$)$^8$B radiative capture. This reaction is important for understanding the solar neutrino flux. Starting from a selected similarity-transformed chiral nucleon-nucleon interaction that accurately describes two-nucleon data, we performed parameter-free many-body calculations that simultaneously predict both the normalization and the shape of the S-factor. We study the dependence on the number of $^7$Be eigenstates included in the coupled-channel equations and on the size of the harmonic oscillator basis used for the expansion of the eigenstates and of the localized parts of the integration kernels. Our S-factor result at zero energy is on the lower side of, but consistent with, the latest evaluation.  
\end{abstract}


\maketitle

The core temperature of the Sun can be determined with high accuracy through measurements of the $^8$B neutrino flux, currently known with a $\sim9\%$ precision~\cite{SNO}. 
An important input in modeling this flux is the $^7$Be(p,$\gamma$)$^8$B reaction~\cite{Adelberger} that constitutes the final step of the nucleosynthetic chain leading to $^8$B. At solar energies this reaction proceeds by external, predominantly nonresonant $E1$, $S$- and $D$-wave capture into the weakly-bound ground state (g.s.) of $^8$B. 
Experimental determinations of the $^7$Be(p,$\gamma$)$^8$B capture include direct measurements with proton beams on $^7$Be targets~\cite{Filippone,Baby,Seattle} as well as indirect measurements through the breakup of a $^8$B projectile into $^7$Be and proton in the Coulomb field of a heavy target~\cite{BBR86,Be7pgamm_exp,GSI}. Theoretical calculations needed to extrapolate the measured S-factor to the astrophysically relevant Gamow energy were performed with several methods: the R-matrix parametrization~\cite{Barker95}, the potential model~\cite{Robertson,Typel97,Davids03}, microscopic cluster models~\cite{DB94,Csoto95,D04} and, recently, also using the {\it ab initio} no-core shell model wave functions for the $^8$B bound state~\cite{NBC06}. The most recent evaluation of the $^7$Be(p,$\gamma$)$^8$B  S-factor at zero energy, $S_{17}(0)$, has a $\sim$10\% error dominated by the uncertainty in theory~\cite{Adelberger}. 

In this Letter, we present the first parameter-free {\it ab initio} many-body calculations of the $^7$Be(p,$\gamma$)$^8$B capture starting from a nucleon-nucleon ($NN$) interaction that describes two-nucleon properties with high accuracy. We apply a recently developed technique that combines {\em ab initio} no-core shell model (NCSM)~\cite{NCSMC12} and resonating-group method (RGM)~\cite{RGM,RGM1} into a new many-body approach~\cite{NCSMRGM,NCSMRGM_IT,NCSMRGM_dalpha} ({\em ab initio} NCSM/RGM) capable of treating bound and scattering states of light nuclei in a unified formalism.  We use, in particular, the orthonormalized NCSM/RGM many-body wave functions given by 
\begin{eqnarray}
|\Psi^{J^\pi T}\rangle &=& \sum_{\nu\nu^\prime} \int dr r^2 \int dr^\prime r^{\prime 2} \, \hat{\mathcal A}_{\nu}|\Phi^{J^\pi T}_{\nu r}\rangle  
\nonumber \\
&&\times \, {\cal N}^{-1/2}_{\nu\nu^\prime}(r,r^\prime)\, \frac{\chi^{J^\pi T}_{\nu'}(r^\prime)}{r^\prime} \, , \label{trial}
\end{eqnarray}
with the inter-cluster antisymmetrizer $\hat{\mathcal A}_{\nu}$, the center-of-mass separation $\vec r_{A-a,a}$, and binary-cluster channel states
\begin{eqnarray}
|\Phi^{J^\pi T}_{\nu r}\rangle &=& \Big [ \big ( \left|A{-}a\, \alpha_1 I_1^{\,\pi_1} T_1\right\rangle \left |a\,\alpha_2 I_2^{\,\pi_2} T_2\right\rangle\big ) ^{(s T)}\nonumber\\
&&\times\,Y_{\ell}\left(\hat r_{A-a,a}\right)\Big ]^{(J^\pi T)}\,\frac{\delta(r-r_{A-a,a})}{rr_{A-a,a}}\,.\label{basis}
\end{eqnarray}
The wave functions $\chi^{J\pi T}_\nu(r)$ of the relative inter-cluster motion satisfy the integro-differential coupled-channel equations
\begin{equation}
{\sum_{\nu^\prime}\!\!\int \!\!dr^\prime r^{\prime\,2}} [{\mathcal N}^{-\frac12}{\mathcal H}\,{\mathcal N}^{-\frac12}]_{\nu\nu^\prime\,}\!(r,r^\prime)\frac{\chi_{\nu^\prime} (r^\prime)}{r^\prime} \!=\! E\,\frac{\chi_{\nu} (r)}{r}  \label{RGMeq}
\end{equation}
with bound- or scattering-state boundary conditions. The Hamiltonian and norm kernels,
\begin{eqnarray}
{\mathcal H}^{J^\pi T}_{\nu^\prime\nu}(r^\prime, r) &=& \left\langle\Phi^{J^\pi T}_{\nu^\prime r^\prime}\right|\hat{\mathcal A}_{\nu^\prime}H\hat{\mathcal A}_{\nu}\left|\Phi^{J^\pi T}_{\nu r}\right\rangle\,,\label{H-kernel}\\
{\mathcal N}^{J^\pi T}_{\nu^\prime\nu}(r^\prime, r) &=& \left\langle\Phi^{J^\pi T}_{\nu^\prime r^\prime}\right|\hat{\mathcal A}_{\nu^\prime}\hat{\mathcal A}_{\nu}\left|\Phi^{J^\pi T}_{\nu r}\right\rangle\,,\label{N-kernel}
\end{eqnarray}
contain all the nuclear structure and antisymmetrization properties of the problem.
Further relevant details of the NCSM/RGM formalism are given in Ref.~\cite{NCSMRGM}. 

\begin{figure}
\begin{minipage}{8cm}
\includegraphics*[width=0.9\columnwidth]{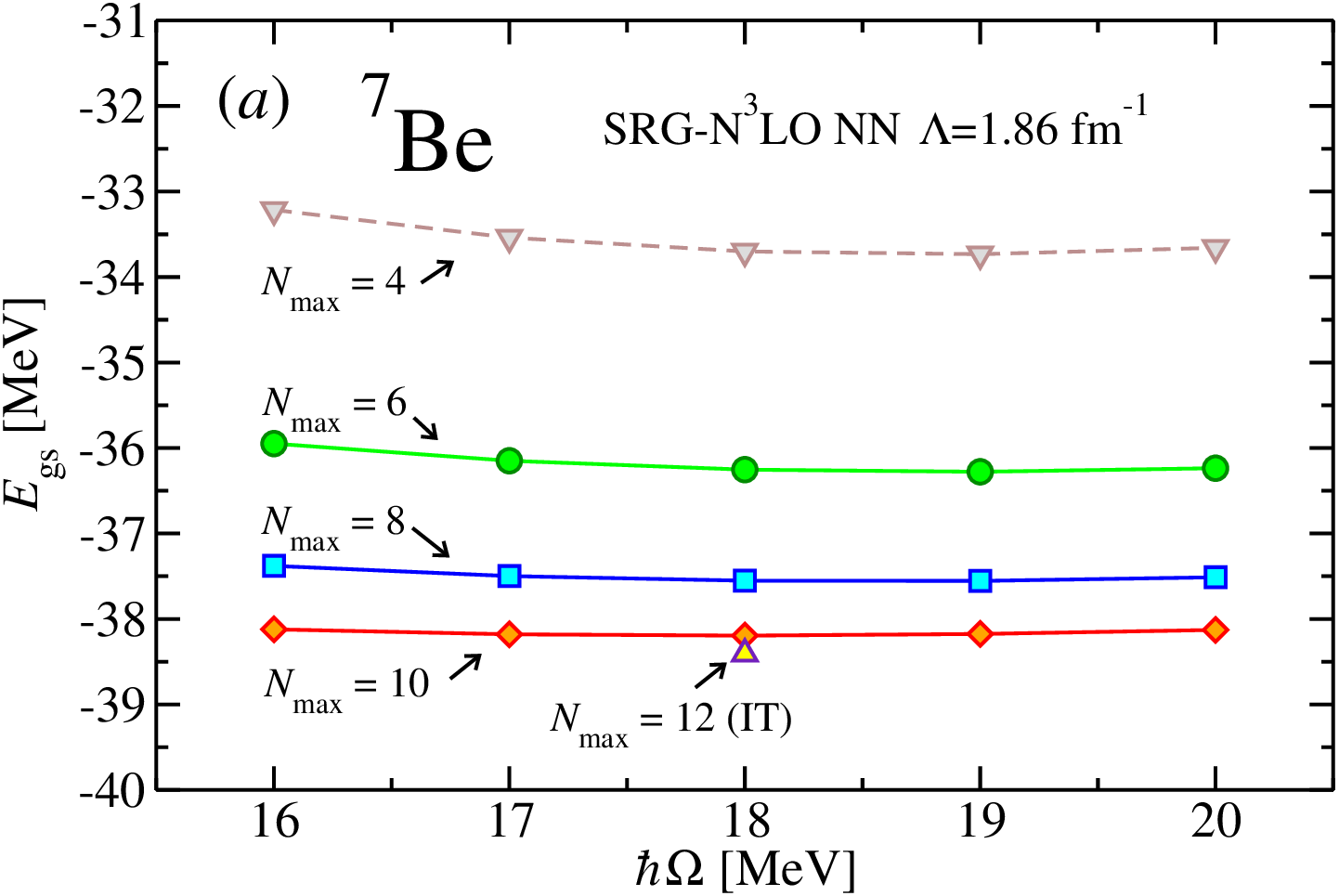}
\end{minipage}
\hfill
\begin{minipage}{8cm}
\includegraphics*[width=0.9\columnwidth]{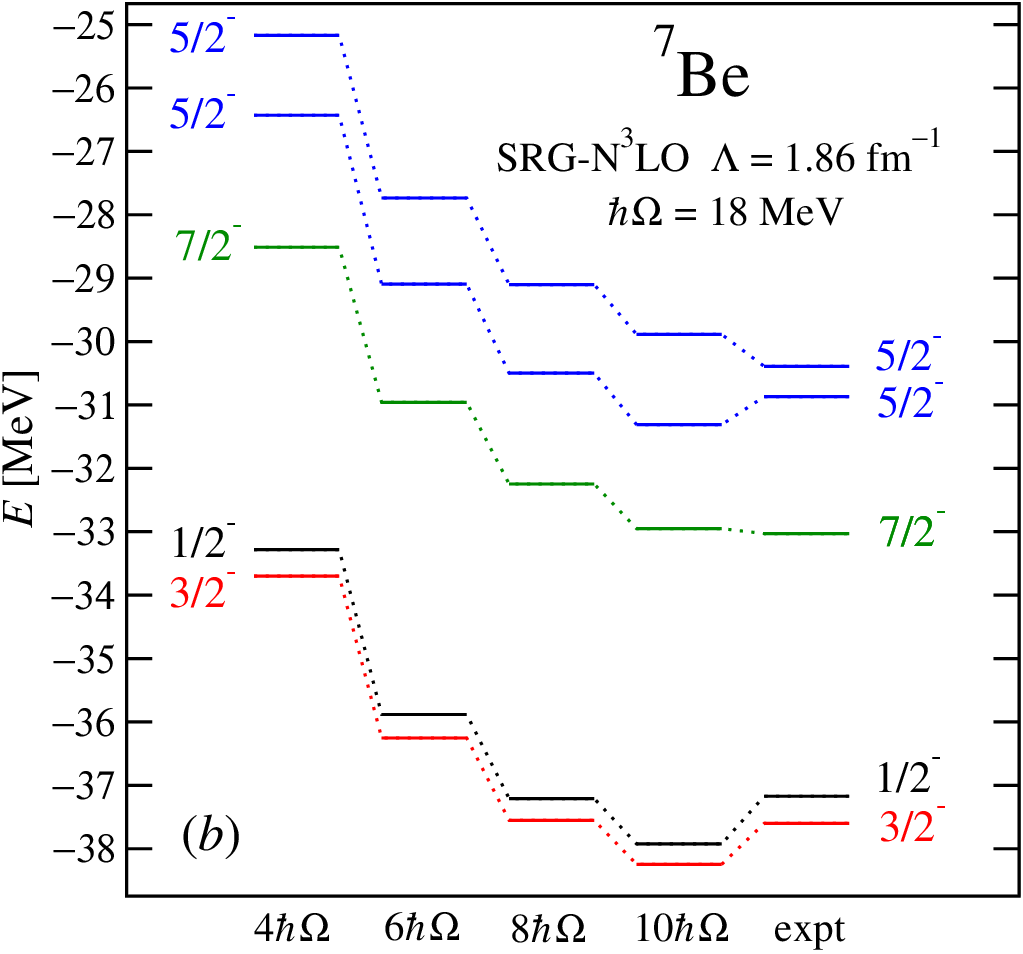}
\end{minipage}
\caption{(Color online) Calculated NCSM $^7$Be eigenenergies using the SRG-N$^3$LO $NN$ potential with $\Lambda=1.86$ fm$^{-1}$. Panel $(a)$ shows the dependence of the g.s.\  energy on the HO frequency for $N_{\rm max}=4{-}12$ (with $N_{\rm max}=12$ results obtained within the importance-truncated basis). Absolute energies of the lowest 5 eigenstates for $N_{\rm max}=4-10$ and $\hbar\Omega=18$ MeV are compared to experimental values in panel $(b)$.}
\label{fig:Be7_energy}
\end{figure}
In the present case $A$ is equal to 8, and the projectile is a proton [$a{=}1$ in Eq.~(\ref{basis})]. The input into Eq.~(\ref{RGMeq}) are: $(i)$ the chiral N$^3$LO $NN$ potential~\cite{N3LO}, which we soften by a similarity renormalization group (SRG) transformation~\cite{SRG,Roth_SRG} characterized by an evolution parameter $\Lambda$; $(ii)$ the eigenstates of the target, {\em i.e.} $^7$Be, calculated within the NCSM. 
In Fig.~\ref{fig:Be7_energy}, we show the energy dependence of the $^7$Be g.s.\ on the harmonic-oscillator (HO) frequency $(a)$ for the HO basis sizes $N_{\rm max}=4$ to 12, with the $12\hbar\Omega$ results obtained using the importance-truncation scheme~\cite{IT-NCSM}. The frequency dependence is quite flat and, with the selected $NN$ potential, we reach converge for 
the g.s.\ at $N_{\rm max}\approx12$. The g.s. energy minimum is found at $\hbar\Omega{=}18$ MeV and we choose this frequency for all subsequent calculations (including eigenstates and integration kernels). The convergence of the absolute energies of the lowest five $^7$Be states is presented in panel $(b)$ of Fig.~\ref{fig:Be7_energy}. Compared to the experimental values, we observe a slight overbinding of the g.s.\  and an overestimation of the $7/2^-$ and $5/2^-_2$ state excitation energies, but, overall, the agreement is reasonable. In Table~\ref{tab:Be7}, we compare some of our (IT-)NCSM $^7$Be results based on calculations up to $N_{\rm max}{=}14$ to experimental values.
\begin{table}
\caption{$^7$Be g.s.\ energy (in MeV), charge radius (in fm), g.s.\ quadrupole (in $e$ fm$^2$) and magnetic (in $\mu_{\rm N}$) moments and M1 transition (in $\mu_{\rm N}^2$) obtained within the NCSM using the SRG-N$^3$LO $NN$ potential with $\Lambda=1.86$ fm$^{-1}$. Experimental values are from Refs.~\protect\cite{A=5-7,Nortershauser}.
\label{tab:Be7}}
\begin{ruledtabular}
\begin{tabular}{cccccc}
 & $E_{g.s.}$ & $r_{\rm c}$ & Q & $\mu$  & B(M1;$\frac{1}{2}^-{\rightarrow}\frac{3}{2}^-$)\\
\hline
NCSM  & -38.46         & 2.46(2)     & -5.39(10)  & -1.15          & 3.14      \\
Expt.   & -37.60         & 2.647(17) &     -           & -1.3995(5)  & 3.71(48) \\
\end{tabular}
\end{ruledtabular}
\end{table}
\begin{figure}
\includegraphics*[width=0.9\columnwidth]{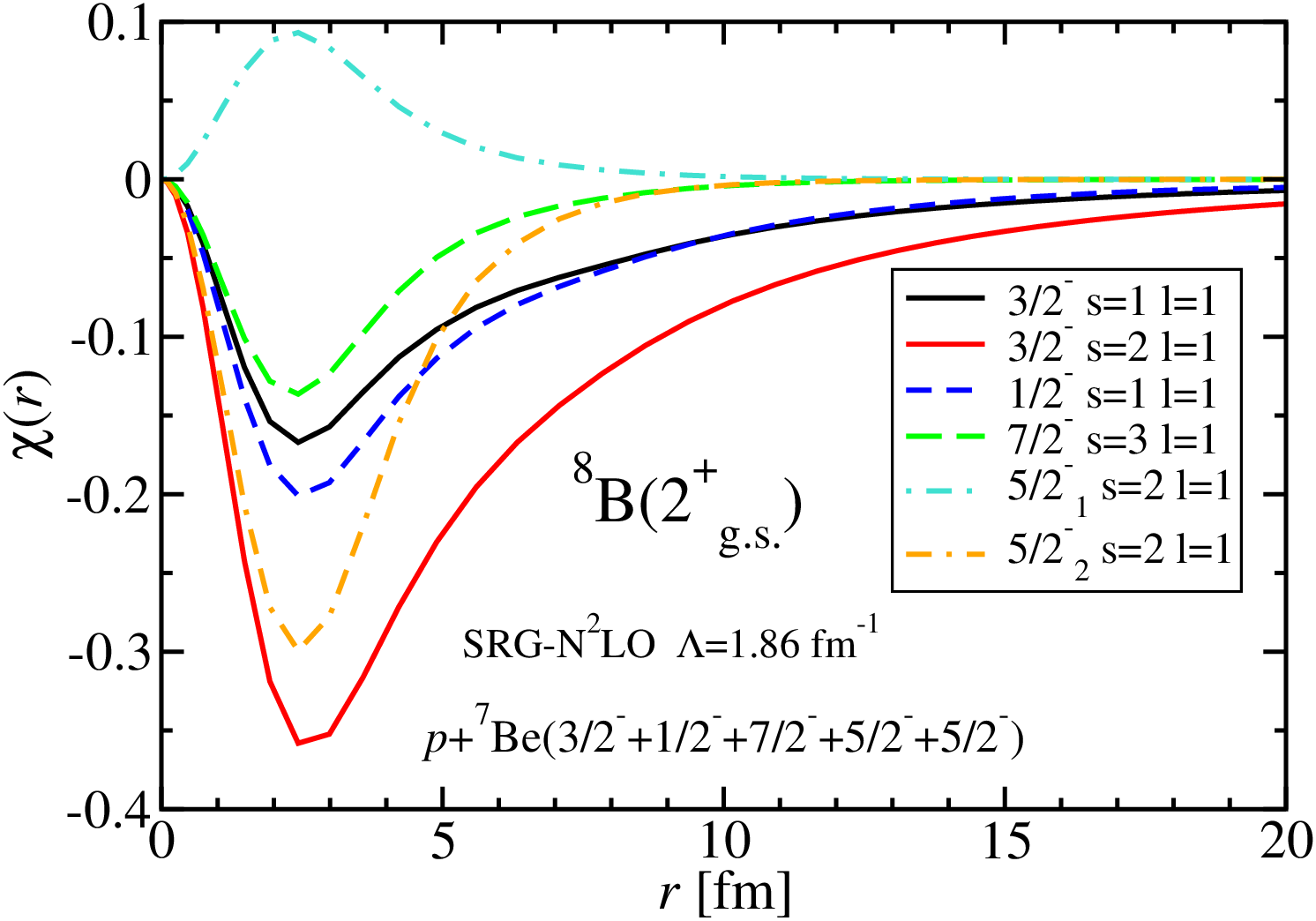}
\caption{(Color online) Dominant $P$-wave components of the $2^+$ $^8$B g.s.\ wave function for $N_{\rm max}=10$ and $\hbar\Omega=18$ MeV, using the SRG-N$^3$LO $NN$ potential with $\Lambda=1.86$ fm$^{-1}$. The NCSM/RGM calculation includes $^7$Be g.s.\ and $1/2^-$, $7/2^-$, $5/2^-_1$ and  $5/2^-_2$ excited states. The calculated s.e.\ is 136 keV.}
\label{fig:B8_gs}
\end{figure}
Using the five lowest $N_{\rm max}{=}10$  eigenstates of $^7$Be, we first solve Eq.~(\ref{RGMeq}) with 
bound-state boundary conditions to find the g.s.\  of $^8$B. We note that the same  $N_{\rm max}$ ($N_{\rm max}$ +1 for the positive parity states) value is used to expand the localized parts of the integrations kernels (\ref{H-kernel}) and (\ref{N-kernel}). The chosen SRG-N$^3$LO $NN$ potential with $\Lambda{=}1.86$ fm$^{-1}$ leads to a single bound state, $2^+$, with separation energy (s.e.) 136 keV, quite close to the experimental 137 keV. For the calculation of the low-energy behavior of the $S_{17}$ S-factor, a correct s.e.\ is very important. The fact that 
the experimental s.e.\ of $^8$B can be found using the SRG potential with a $\Lambda$ from a``natural'' range, i.e. ${\approx} 1.8 {-} 2.1$~fm$^{-1}$, is reassuring.  In Fig.~\ref{fig:B8_gs}, we plot the most significant components of the radial wave functions $\chi(r)$ for the $2^+$ g.s. of $^8$B. The dominant component is clearly the channel-spin $s{=}2$ $P$-wave of the $^7$Be(g.s.)-$p$ that extends to a distance far beyond the plotted range. Remarkably, we notice a substantial contribution from the $^7$Be($5/2^-_2$)-$p$ $P$-wave. Clearly, for a realistic description of the $^8$B g.s., this state must be taken into account. The influence of still higher $^7$Be resonances on the S-factor results will be discussed at the end of this Letter. 

\begin{figure}
\includegraphics*[width=0.8\columnwidth]{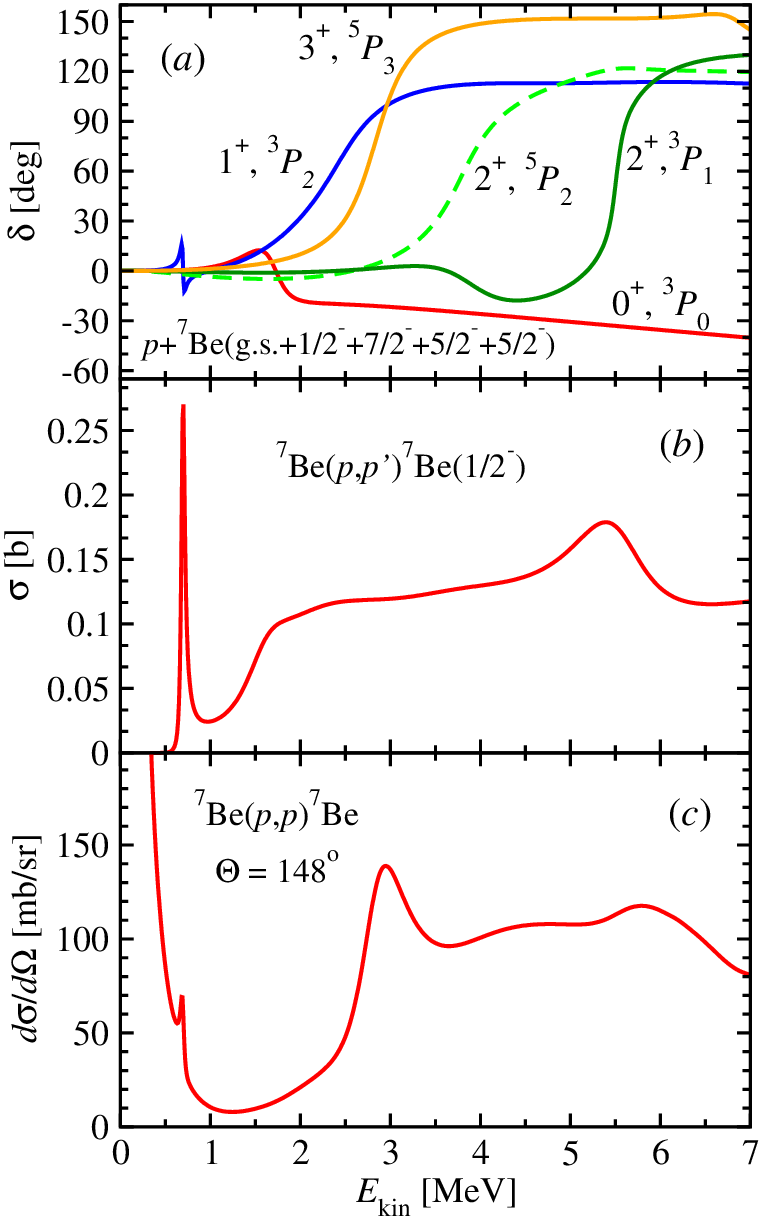}
\caption{(Color online) Selected $P$-wave diagonal phase shifts of $p$-$^7$Be elastic scattering $(a)$, inelastic $^7$Be($p$,$p'$)$^7$Be(1/2$^-$) cross section $(b)$ and elastic $^7$Be($p$,$p$)$^7$Be  differential cross section at $\Theta_{c.m.}=148^0$ $(c)$. Calculations as described in Fig.~\protect\ref{fig:B8_gs}.}
\label{fig:Be7_31755_Pwaves}
\end{figure}
Next, we solve the same NCSM/RGM equations~(\ref{RGMeq}) with scattering-state boundary conditions for a chosen range of energies and obtain scattering wave functions and the scattering matrix. The resulting phase shifts and cross sections are displayed in Fig.~\ref{fig:Be7_31755_Pwaves}. All energies are in the center of mass (c.m.). We find several $P$-wave resonances in the considered energy range. The first $1^+$ resonance, manifested in both the elastic and inelastic cross sections, corresponds to the experimental $^8$B $1^+$ state at $E_x{=}0.77$ MeV (0.63 MeV above the $p$-$^7$Be threshold)~\cite{TUNL_A8}. The $3^+$ resonance, responsible for the peak in the elastic cross section, corresponds to the experimental $^8$B $3^+$ state at $E_x{=}2.32$ MeV. However, we also find a low-lying $0^+$ and additional $1^+$ and $2^+$ resonances that can be distinguished in the inelastic cross section. In particular, the $s{=}1$ $P$-wave $2^+$ resonance is clearly visible. There is also an $s{=}2$ $P$-wave $2^+$ resonance with some impact on the elastic cross section. These resonances are not included in the current $A{=}8$ evaluation~\cite{TUNL_A8}. We note, however, that the authors of the recent Ref.~\cite{Mitchell10} do claim observation of  low-lying $0^+$ and $2^+$ resonances based on an R-matrix analysis of their $p$-$^7$Be scattering experiment. Their suggested $0^+$ resonance at 1.9 MeV is quite close to the calculated $0^+$ energy of the present work.

\begin{figure}
\begin{minipage}{8cm}
\includegraphics*[width=0.9\columnwidth]{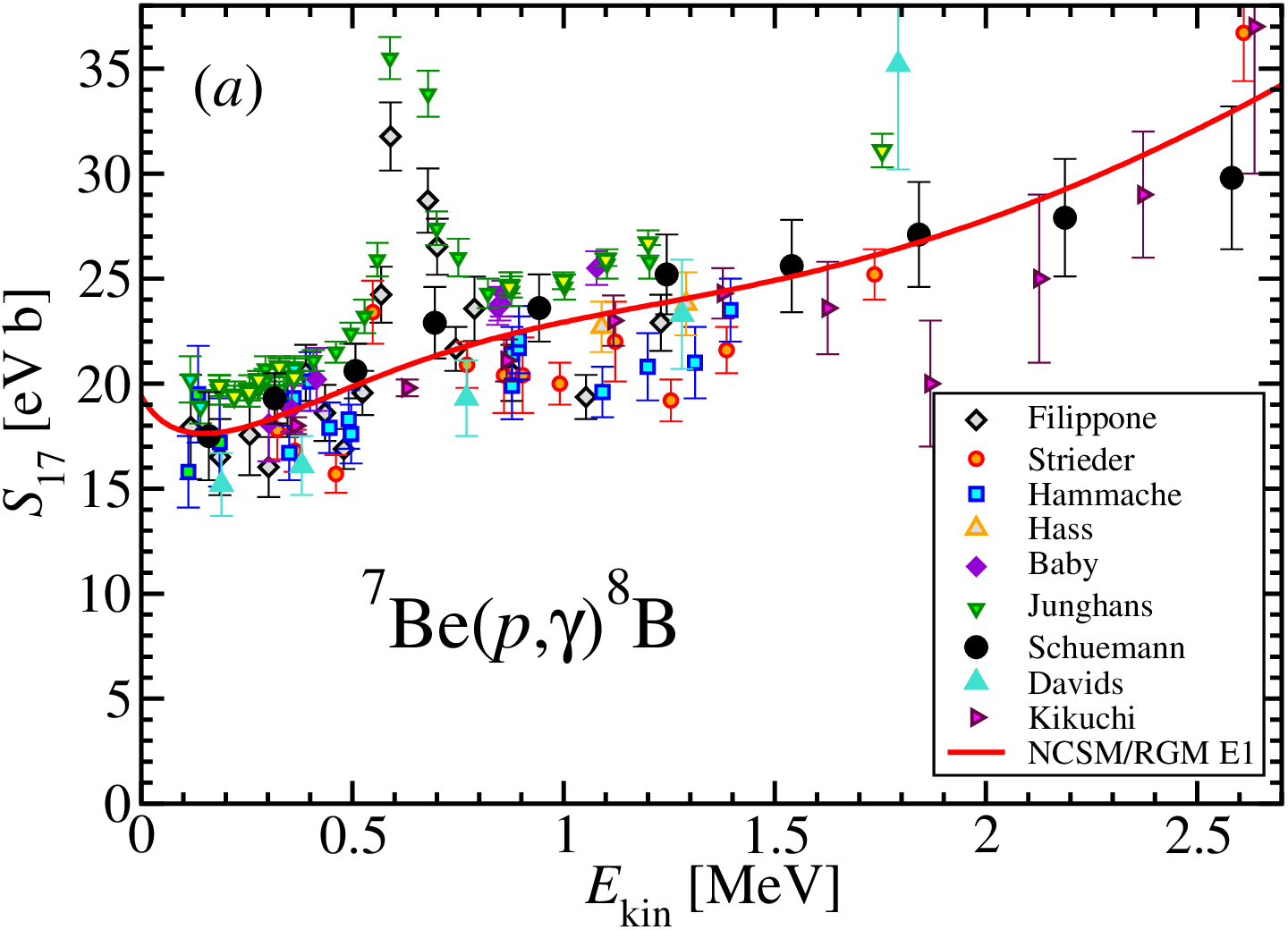}
\end{minipage}
\hfill
\begin{minipage}{8cm}
\includegraphics*[width=0.9\columnwidth]{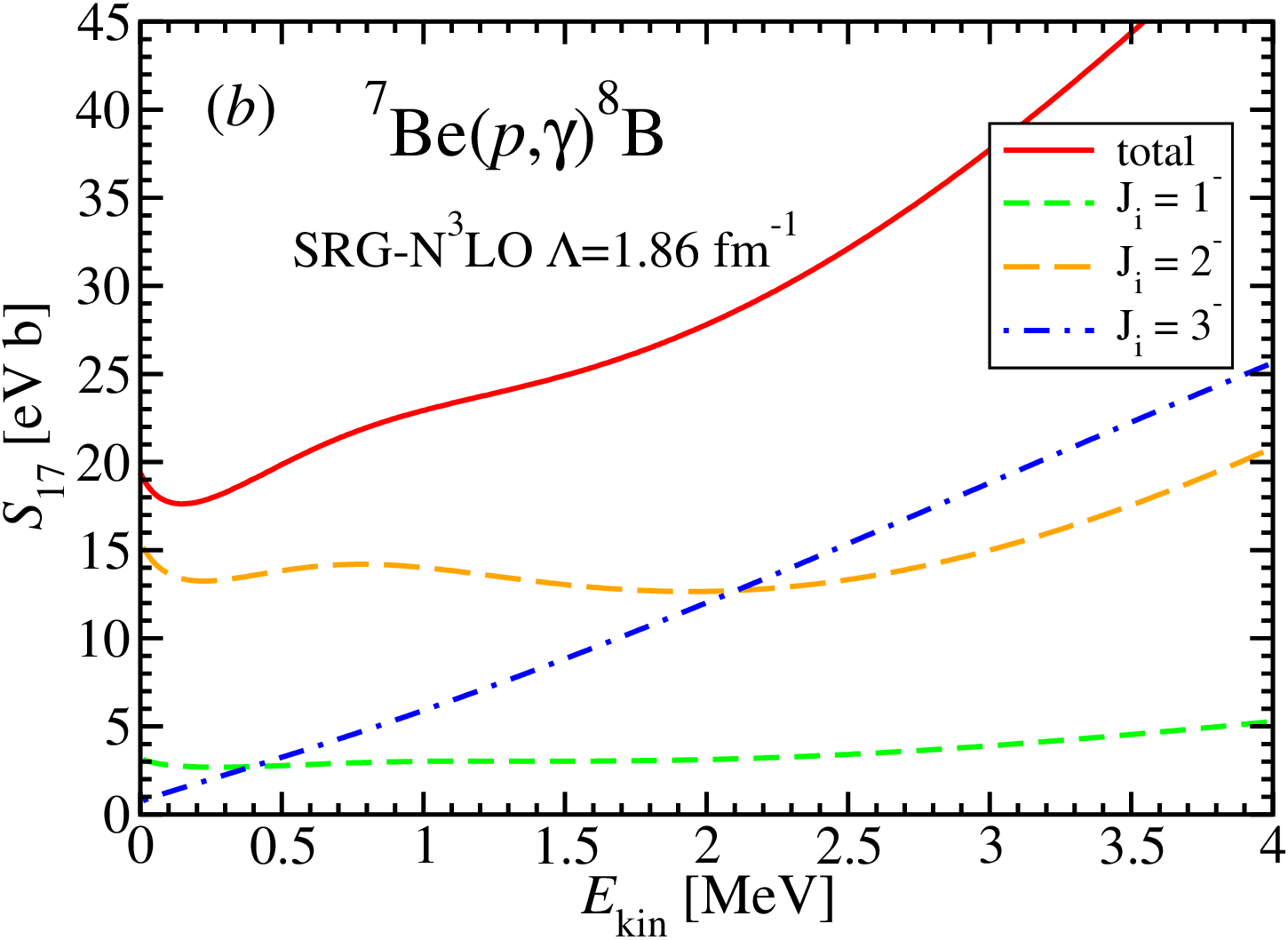}
\end{minipage}
\caption{(Color online) Calculated $^7$Be($p$,$\gamma$)$^8$B S-factor as function of the energy in the c.m.\ compared to data $(a)$. Only $E1$ transitions were considered. Initial-state partial wave contributions are shown in panel $(b)$. Calculation as in Fig.~\protect\ref{fig:B8_gs}.}
\label{fig:Be7_p_g_1p86} 
\end{figure}
With the resulting bound- and scattering-state wave functions that are properly orthonormalized and antisymmetrized~(\ref{trial}), we calculate the $^7$Be($p$,$\gamma$)$^8$B radiative capture using a one-body $E1$ transition operator. The resulting $S_{17}$ factor is compared to several experimental data sets in panel $(a)$ of Fig.~\ref{fig:Be7_p_g_1p86}.  In the data, one can see also the contribution from the $1^+$ resonance due to the $M1$ capture.  Our calculated S-factor is somewhat lower than the recent Junghans data~\cite{Seattle} but the shape reproduces closely the trend of the GSI data~\cite{GSI} and is quite similar to that obtained within the microscopic cluster model~\cite{D04} used in the most recent $S_{17}$ evaluation~\cite{Adelberger}. The contributions from the initial $1^-$, $2^-$ and $3^-$ partial waves are shown in panel $(b)$ of Fig.~\ref{fig:Be7_p_g_1p86}. Our calculated $S_{17}(0){\approx} 19.4$ eV b is on the lower side, but consistent with the latest evaluation 20.8$\pm 0.7$(expt)$\pm1.4$(theory) eV b.

\begin{figure}
\includegraphics*[width=0.8\columnwidth]{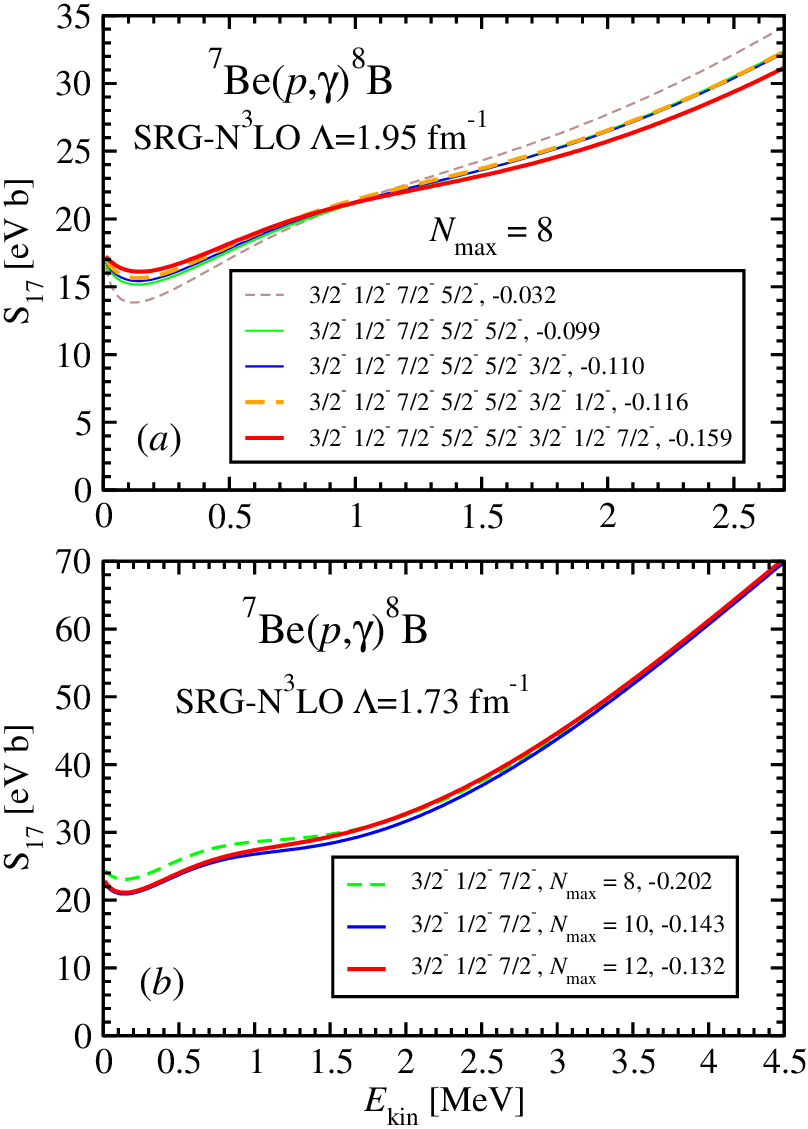}
\caption{(Color online) Convergence of the $^7$Be($p$,$\gamma$)$^8$B S-factor with the number of $^7$Be eigenstates $(a)$ and the size of the HO basis used to expand the $^7$Be eigenstates and localized parts of the integration kernels $(b)$. The number of eigenstates and the calculated separation energy in each case is shown in the legend. HO frequencies of $\hbar\Omega=19$ MeV $(a)$ and 17 MeV $(b)$ corresponding to the respective minima of $^7$Be g.s. were used.}
\label{fig:Be7_p_g_conv} 
\end{figure}
We studied the convergence of the $^7$Be NCSM calculations in Fig.~\ref{fig:Be7_energy}. To verify the behavior of our S-factor with respect to HO basis size and number of included $^7$Be eigenstates, we performed additional calculations as summarized in Fig.~\ref{fig:Be7_p_g_conv}. To study the dependence on the HO basis size, shown up to $N_{\rm max}{=}12$ in panel $(b)$, we use the importance truncation scheme and, due to computational limitations, we include only the three lowest eigenstates of $^7$Be. The $N_{\rm max}{=}10$ and 12 S-factors are very close. In panel $(a)$, we present results with up to 8 $^7$Be eigenstates obtained in a $N_{\rm max}{=}8$ basis. Calculations with more than 5 eigenstates are presently out of reach for larger $N_{\rm max}$ values. We can see a significant impact of the $5/2^-$ states (with only three $^7$Be states, $^8$B is unbound in this case). Among the others only the 8th state, $7/2^-_2$, contributes somewhat to the s.e.\ and flattens the S-factor at higher energies. We note that we selected different SRG-N$^3$LO $NN$ potentials with the aim to match closely the experimental s.e.\ in each of the largest calculation. From these results we conclude that the use of the $N_{\rm max}{=}10$ space is justified and a limitation to the five lowest $^7$Be eigenstates is quite reasonable (or that the $N_{\rm max}{=}8$ space is insufficient and a limitation to just 3 states is unrealistic). Also, based on these results we estimate the uncertainty of our calculated $S_{17}(0)$ to be $\pm 0.7$ eV b.

In conclusion, we performed parameter-free {\it ab initio} many-body calculations of the $^7$Be($p$,$\gamma$)$^8$B radiative capture that predict simultaneously both the normalization and the shape of the S-factor. Our S-factor result at zero energy, $S_{17}(0){=}19.4(7)$ eV b, is on the lower side of, but consistent with, the latest evaluation, and its shape follows closely the data from Ref.~\cite{GSI}. Our calculations can be further improved by including effects of additional higher-lying $^7$Be resonances. This can be best done by coupling the NCSM/RGM binary-cluster basis with the NCSM calculations for $^8$B as outlined in Ref.~\cite{NCSM_review}. The inclusion of three-nucleon interactions, both chiral and SRG-induced~\cite{JNF09}, is also desirable. Efforts in these directions are under way. 

Computing support for this work came from the LLNL Institutional Computing Grand Challenge program and the J\"ulich Supercomputing Centre. Prepared in part by LLNL under Contract DE-AC52-07NA27344. Support from the NSERC grant No. 401945-2011, from the UNEDF SCIDAC DOE Grant DE-FC02-07ER41457, the Deutsche Forschungsgemeinschaft through contract SFB 634, and the Helmholtz International Center for FAIR (HIC for FAIR) is acknowledged.


\begin{thebibliography}{10}

\bibitem{SNO} SNO Collaboration, S. N. Ahmed {\it et al.}, Phys. Rev. Lett. {\bf 92},
              181301 (2004).

\bibitem{Adelberger} E. G. Adelberger {\it et al.}, Rev. Mod. Phys. {\bf 83}, 195 (2011); Rev. Mod. Phys. {\bf 70}, 1265 (1998).

\bibitem{Filippone} B. W. Filippone, A. J. Elwyn, C. N. Davids, and D. D. Koetke,
                    Phys. Rev. Lett {\bf 50}, 412 (1983);
                    Phys. Rev. C {\bf 28}, 2222 (1983).

\bibitem{Baby}
   L.T. Baby {\it et al.}, Phys. Rev. Lett. {\bf 90}, 022501 (2003).

\bibitem{Seattle} A. R. Junghans {\it et al.}, 
 Phys. Rev. C  {\bf 68}, 065803 (2003).

\bibitem{BBR86}
   G.\ Baur, C.\ A.\ Bertulani and H.\ Rebel,  Nucl.\  Phys.\   {\bf A458}, 188 (1986).\ 

\bibitem{Be7pgamm_exp}  
   N.\  Iwasa {\it et al.}, Phys.\ Rev.\ Lett.\ {\bf 83}, 2910 (1999);
   B.\ Davids {\it et al.}, Phys.\  Rev.\  Lett.\  {\bf 86}, 2750 (2001).

\bibitem{GSI} F.\  Sch\"{u}mann {\it et al.}, Phys.\  Rev.\  C {\bf 73},015806 (2006);
Phys.\ Rev.\ Lett.\ {\bf 90}, 232501 (2003).

\bibitem{Barker95} F.\ C.\ Barker, Nucl.\  Phys.\  {\bf A588}, 693 (1995).

\bibitem{Robertson} R.\  G.\  H.\  Robertson, Phys.\  Rev.\  C {\bf 7}, 543 (1973).

\bibitem{Typel97} S.\  Typel, H.\  H.\  Wolter, and G.\  Baur, Nucl.\  Phys.\  {\bf A613},
                  147 (1997).

\bibitem{Davids03} B.\  Davids and S.\  Typel, Phys.\  Rev.\  C {\bf 68}, 045802 (2003).

\bibitem{DB94} P.\  Descouvemont and D.\  Baye, Nucl.\  Phys.\  {\bf A567}, 341 (1994).

\bibitem{Csoto95} A.\  Cs\'ot\'o, K.\  Langanke, S.\  E.\  Koonin, and T.\  D.\  Shoppa,
                  Phys.\ Rev.\  C {\bf 52}, 1130 (1995).

\bibitem{D04} P.\ Descouvemont, Phys.\ Rev.\ C {\bf 70}, 065802 (2004).

\bibitem{NBC06} P.\ Navratil, C.\ A.\ Bertulani and E.\ Caurier, Phys.\ Lett B {\bf 634}, 191 (2006); Phys.\ Rev.\ C {\bf 73}, 065801 (2006).

\bibitem{NCSMC12} P.\ Navr\'atil, J.\ P.\ Vary, and B.\ R.\ Barrett,
                   Phys.\ Rev.\ Lett. {\bf 84}, 5728 (2000);
                   Phys.\ Rev.\ C {\bf 62}, 054311 (2000).

\bibitem{RGM} K.\ Wildermuth and Y.\ C.\ Tang, {\it A unified theory of the nucleus},
              (Vieweg, Braunschweig, 1977). 

\bibitem{RGM1} Y.\ C.\ Tang, M. LeMere and D. R. Thompson,
              Phys.\ Rep.\ {\bf 47}, 167 (1978).
             
\bibitem{NCSMRGM} S.\ Quaglioni and P.\ Navr{\'a}til,  Phys.\ Rev.\ Lett.\ {\bf 101}, 092501 (2008);
Phys.\ Rev.\ C  {\bf 79}, 044606 (2009).     


\bibitem{NCSMRGM_IT} P.\ Navr{\'a}til, R.\ Roth and S.\ Quaglioni, Phys.\ Rev.\ C  {\bf 82}, 034609 (2010).
           
\bibitem{NCSMRGM_dalpha} P.\ Navr{\'a}til and S.\ Quaglioni, Phys.\ Rev.\ C  {\bf 83}, 044609 (2011).
 
\bibitem{N3LO} D.\ R.\ Entem and R.\ Machleidt, Phys.\ Rev.\ C {\bf 68}, 041001(R) (2003).
        
\bibitem{SRG} S.\ K.\ Bogner, R.\ J.\ Furnstahl and R.\ J.\ Perry, Phys.\ Rev.\ C {\bf 75}, 061001 (2007).

\bibitem{Roth_SRG} R.\ Roth, S.\ Reinhardt and H.\ Hergert, Phys.\ Rev.\ C {\bf 77}, 064003 (2008);
R. Roth, T. Neff, and H. Feldmeier, Prog. Part. Nucl. Phys. 65, 50 (2010). 
 
\bibitem{IT-NCSM} R.\ Roth and P.\ Navr\'atil, Phys.\ Rev.\ Lett.\ {\bf 99}, 092501 (2007);
R.\ Roth, Phys.\ Rev.\ C {\bf 79}, 064324 (2009).

\bibitem{A=5-7} D.\ R.\ Tilley {\it et al.}, 
            Nucl.\ Phys.\ {\bf A 708}, 3 (2002).

\bibitem{Nortershauser} W.\ N\"ortersh\"auser {\it et al.}, Phys.\ Rev.\ Lett.\ {\bf 102}, 062503 (2009).

\bibitem{TUNL_A8}  D.\ R.\ Tilley {\it et al.}, Nuclear Physics A {\bf 745}, 155 (2004).

\bibitem{Mitchell10}  J.\ P.\ Mitchell {\it et al.}, 
Phys.\ Rev.\ C {\bf 82}, 011601(R) (2010).

\bibitem{NCSM_review} P.\ Navratil, S.\ Quaglioni, I.\ Stetcu and B.\ R.\ Barrett, J.\ Phys.\ G: Nucl.\ Part.\ Phys.\ {\bf 36}, 083101 (2009).

\bibitem{JNF09} E.\ D.\ Jurgenson, P.\ Navratil, and R.\ J.\ Furnstahl, Phys.\ Rev.\ Lett.\ {\bf 103}, 082501 (2009).




\end{thebibliography}
\end{document}